\documentclass{PoS}

\newcommand{\mcprod}[1]{\texttt{prod#1}}

\title{Baseline telescope layouts of the Cherenkov Telescope Array}

\ShortTitle{Baseline telescope layouts of the Cherenkov Telescope Array}

\author{\speaker{P. Cumani}$^{1}$\thanks{corresponding author}, T. Hassan$^{1 \dagger}$, L. Arrabito$^{2}$, K. Bernl{\"o}hr$^{3}$, J. Bregeon$^{2}$, G. Maier$^{4}$, A. Moralejo$^{1}$ for the CTA Consortium\footnote{Full consortium list at http://cta-observatory.org}\\
       $^{1}$Institut de Fisica d'Altes Energies (IFAE), The Barcelona Institute of Science and Technology, Campus UAB, 08193 Bellaterra (Barcelona) Spain\\
       $^{2}$Laboratoire Univers et Particules de Montpellier - UMR5299, Universit\'e de Montpellier - CNRS/IN2P3, Place Eug\`ene Bataillon - CC 72, 34095 Montpellier C\'edex 05 France\\
       $^{3}$Max-Planck-Institut f{\"u}r Kernphysik, P.O. Box 103980, 
        D-69029 Heidelberg, Germany\\
        $^{4}$ DESY, Platanenallee 6, D-15738 Zeuthen, Germany\\
       E-mail: \email{pcumani@ifae.es}, \email{thassan@ifae.es}}

\abstract{The Cherenkov Telescope Array (CTA) will be the next generation of ground-based instrument for Very High Energy gamma-ray astronomy. It will improve the sensitivity of current telescopes by up to an order of magnitude and provide energy coverage from 20 GeV up to 300 TeV. This improvement will be achieved using a total of 19 and 99 telescopes of three different sizes spread out over 0.4 and 4.5 km$^2$ at two sites, respectively, in the Northern and Southern Hemispheres. After a concerted effort involving three different large-scale Monte Carlo productions performed during the last years, here, the baseline layouts for both CTA sites that should emerge after several years of construction are presented.}

\FullConference{35th International Cosmic Ray Conference --- ICRC2017\\
		10--20 July, 2017\\
		Bexco, Busan, Korea}

\begin{document}

\section{Introduction}
\label{sec:intro}
In a few years, the next generation Very High Energy (VHE) Imaging Atmospheric Cherenkov Telescope (IACT) will start operating. The Cherenkov Telescope Array (CTA) will be the first IACT operating as an open observatory, and will be composed of two arrays positioned in the two hemispheres. In the Northern Hemisphere, 4 Large-Sized Telescopes (LSTs) \cite{LSTgamma} and 15 Medium-Sized Telescopes (MSTs) \cite{MSTgamma, SCTgamma} will be positioned in La Palma (Spain), while in the Southern Hemisphere,  4 LSTs, 25 MSTs, and 70 Small-Sized Telescopes (SSTs) \cite{Montaruli:2015} will be installed on the Paranal site (Chile). This configuration will provide an unprecedented energy coverage from 20 GeV up to 300 TeV and an improved sensitivity with respect to the current generation of instruments by up to an order of magnitude.\\
Three large-scale Monte Carlo (MC) productions were performed in order to obtain the best estimation of CTA future capabilities \cite{APP_CTA_MC, MC_ICRC:2013,Hassan-2015}, assess the influence of the selected construction site on performance \cite{Hassan2017Apj, MC_ICRC_site:2015} and gauge capabilities of different telescope types \cite{Wood:2016, GCT}. Two important conclusions came out of the first two studies. First, telescope layouts focusing either on low or high energies would be disfavoured, while balanced arrays with $\sim$ 4 LSTs, 24 MSTs and 70 SSTs were the preferred option. Second, sites at a not too high altitude ($\sim$2000 m) were to give best overall performances.


The main goal of the third MC production (\mcprod{3}) is to find the optimal layout for a given (baseline) number of telescopes for each selected site, taking into account the following set of considerations (in increasing order of priority):

\begin{enumerate}
    \item Full system performance requirements
    \item Sub-system performance requirements
    \item Administrative and geological constraints of the selected sites
    \item Shadowing between telescopes
    \item Staging, i.e. the available arrays during the construction phase 
    \item Impact on calibration and systematic uncertainties
\end{enumerate}

In the following sections we sum up the knowledge attained from these MC productions and conclude with the proposed final CTA baseline arrays.

\section{Third large-scale MC production and analysis}

\begin{figure}[ht]
\begin{center}
\includegraphics[width=0.3\textwidth]{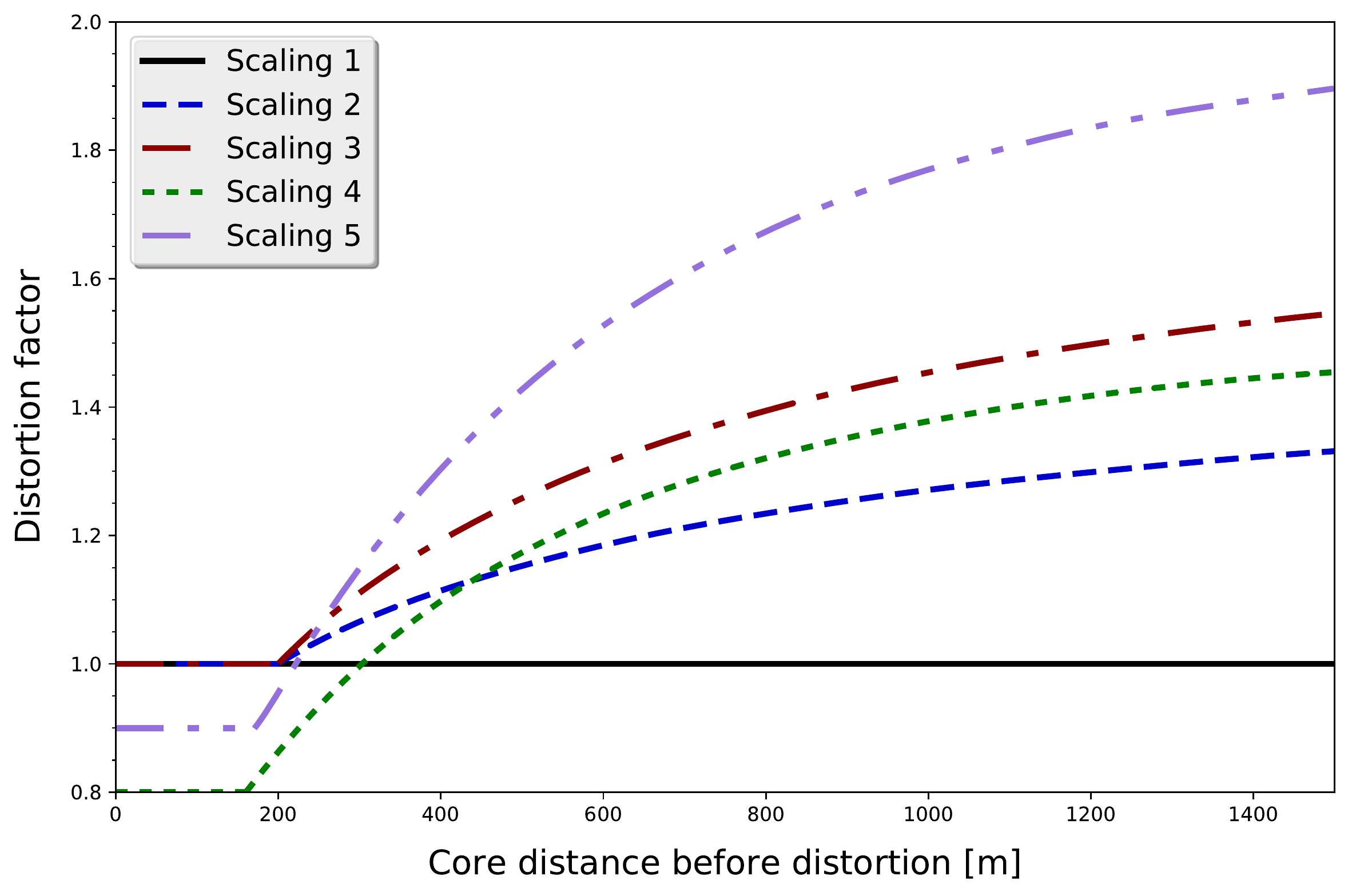}
\includegraphics[width=0.3\textwidth]{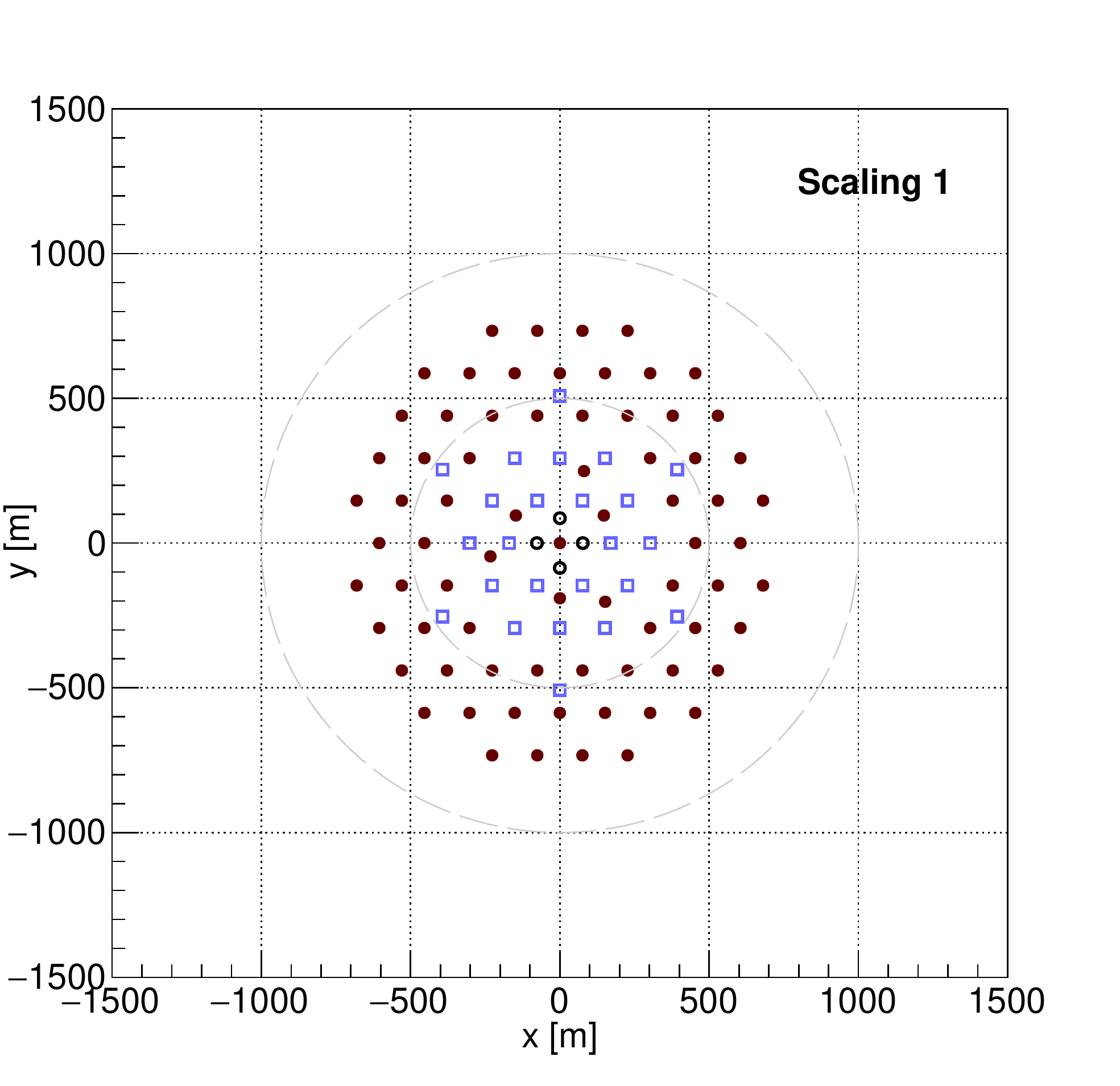}
\includegraphics[width=0.3\textwidth]{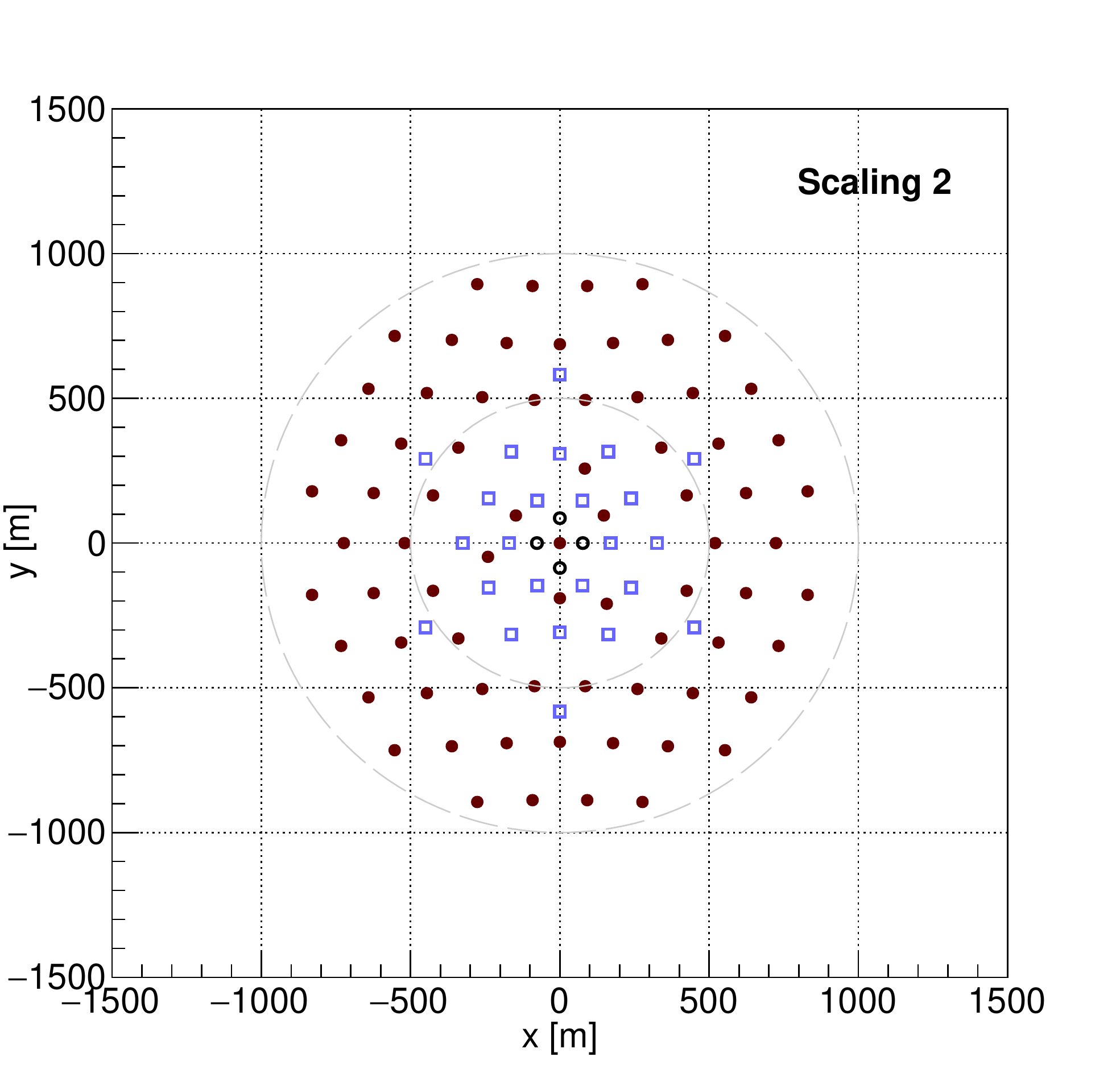}\\
\includegraphics[width=0.3\textwidth]{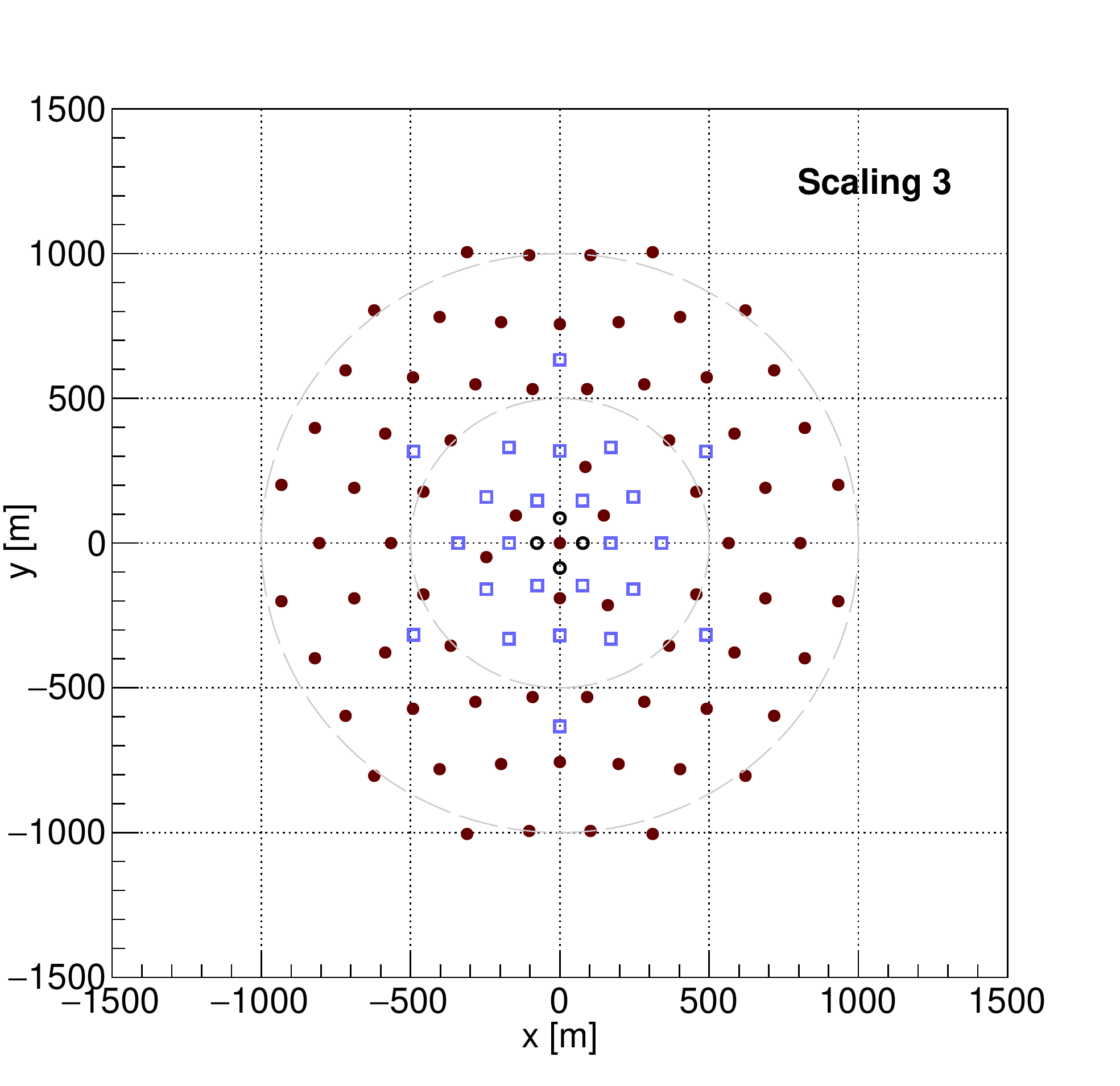}
\includegraphics[width=0.3\textwidth]{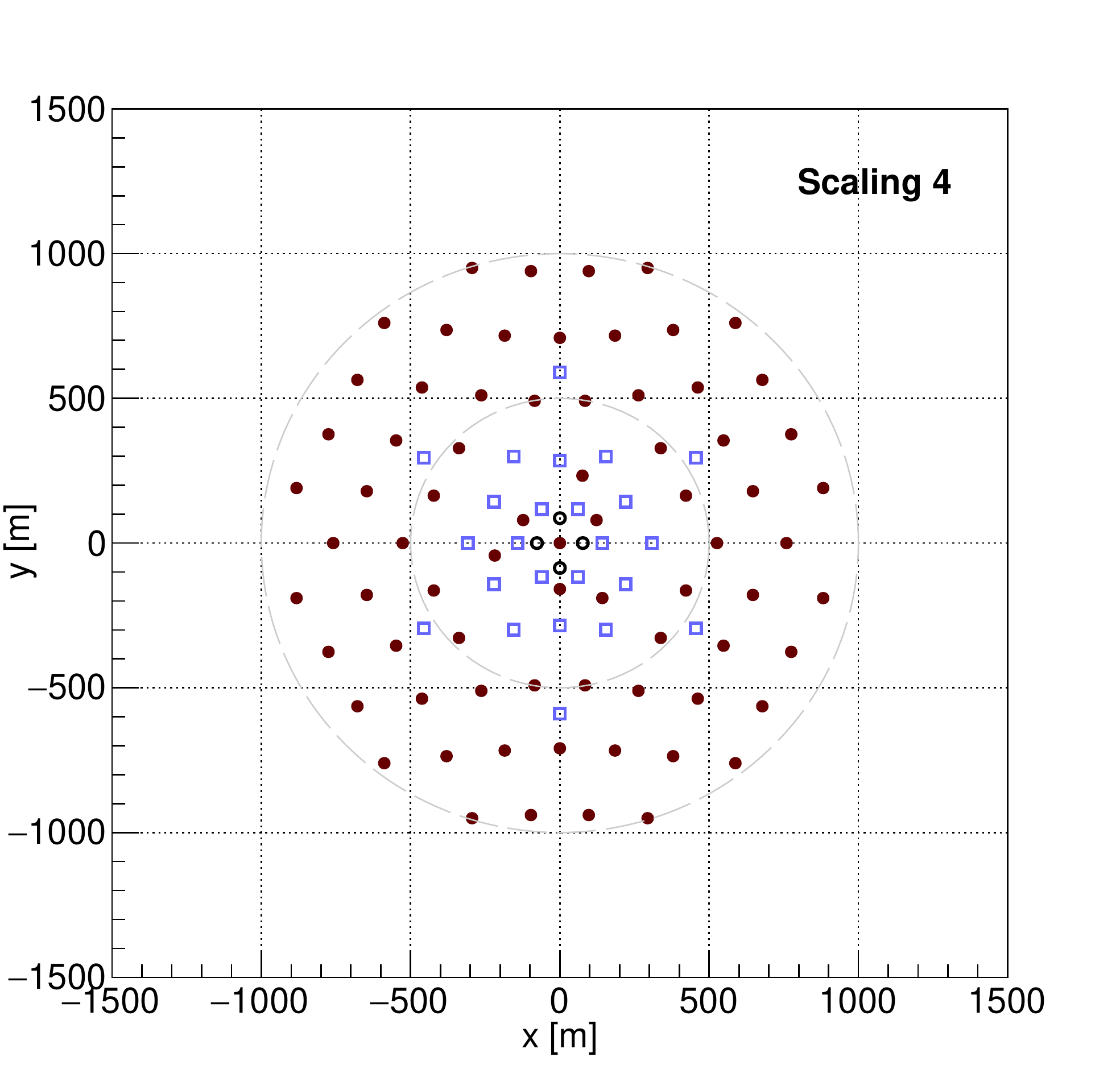}
\includegraphics[width=0.3\textwidth]{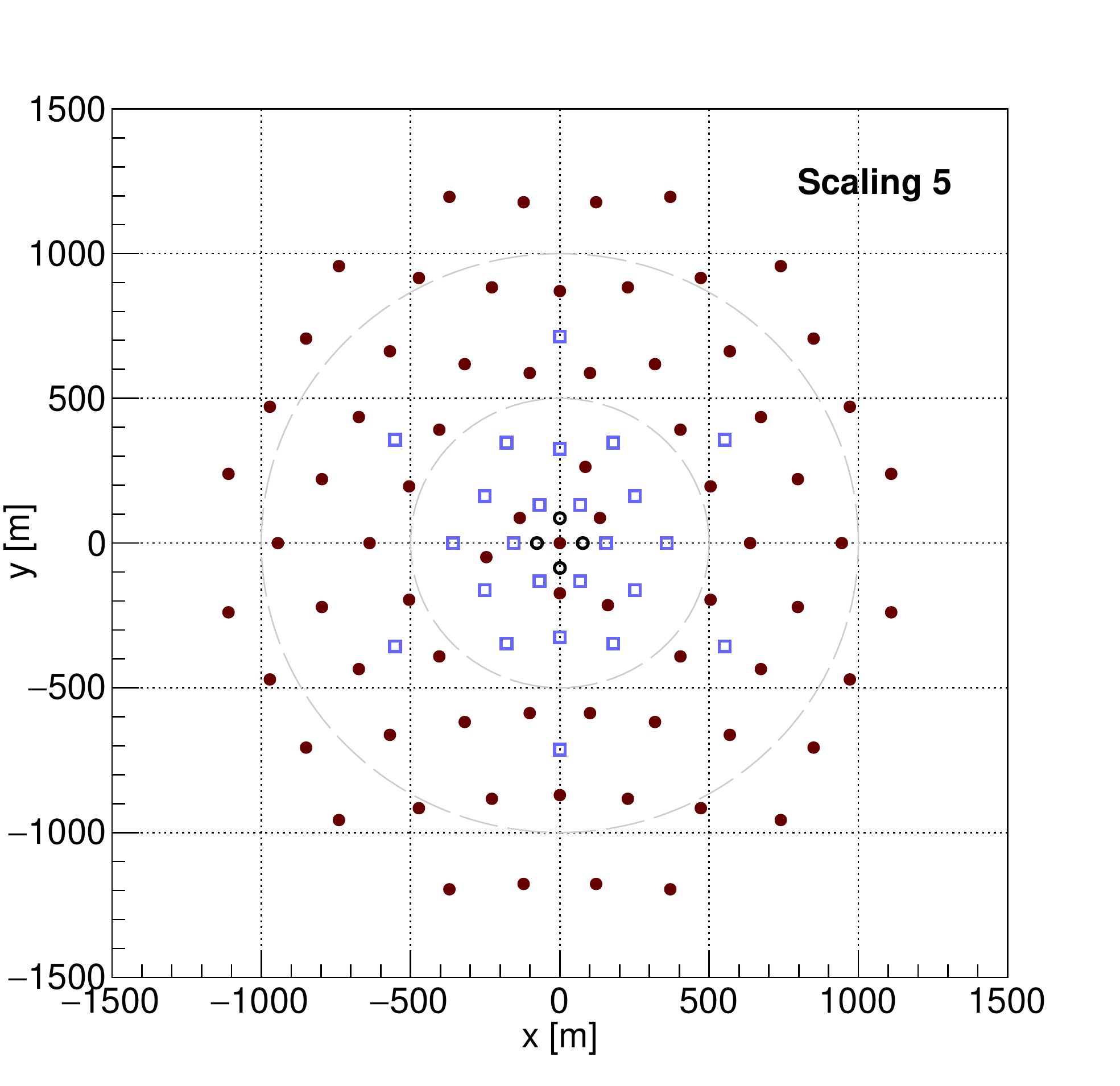}\\
\caption{\textit{Top-left}: Radially symmetric distortion factors for the five different scalings in \mcprod{3} CTA South layouts, as a function of the core distance before this distortion. \textit{Top-right to bottom-right}: an example of the scalings for the Paranal site for one of the studied layouts. LSTs positions are shown with black circles, MSTs with blue squares, and SSTs with filled red circles.}
\label{fig:distfactor}
\end{center}
\end{figure}

Following a parallel approach as in previous MC productions \cite{APP_CTA_MC, Hassan2017Apj}, a dense master layout with many hundreds of telescopes is defined. By selecting sub-samples of telescopes of approximately total equal cost from the master layout and evaluating their performance, conclusions on the most cost-effective CTA array are inferred. The MC production and analysis was mostly run by using the resources of the CTA Computing Grid, and required the use of around 120 million HS06 CPU hours (1200 CPU years) and about 1.4 PBytes of disk storage \cite{Arrabito-2015}.\\

The master layout was produced following these steps:

\begin{itemize}
    \item Initially, an hexagonal telescope grid was defined, from which several plausible baseline layouts can be extracted (see as an example Fig. \ref{fig:distfactor}). In the case of CTA-North (CTA-N), La Palma orography, buildings and streets impose additional constrains on the allowed telescope positions, therefore a custom setup is defined.
    \item The layout is compressed in the East-West direction by a factor $1/ \sqrt{1.12} \approx 0.94$ and stretched in the North-South direction by a factor $\sqrt{1.12} \approx 1.06$. The purpose of this step is to have a more radially symmetric distribution of telescopes in the shower plane projection, following the typical observed zenith angle around culmination.
    \item A set of 5 different radially symmetric (with respect to the center of the array) scaling factors was applied to all telescopes outside of an un-distorted central circle (excluding the LSTs). These homothetic transformations follow the profiles illustrated in Figure \ref{fig:distfactor}. Because of the same topography conditions, the scaled positions for CTA-N follow a slightly different profile.
\end{itemize}

The final master layout used is made up of the combination of these 5 scaled layouts. Apart from the LSTs, the simulations included the two types of MST camera (FlashCam or NectarCam) and three SST types (ASTRI, GCT and SST-1m), leading to a total of more than 3000 simulated telescopes positions. The obtained results do not depend on the particular choice of MST camera or SST type. Some telescope positions, not simulated in the \mcprod{3}, were later included, together with the most promising layouts, in a \mcprod{3b} in order to simulate the final array layouts. There were no differences in the detector models between \mcprod{3} and \mcprod{3b}.

As in \cite{APP_CTA_MC, Hassan2017Apj}, performances were evaluated at 20$^\circ$ zenith angle and, to take into account the effect of the geomagnetic field, two different pointing directions in azimuth (pointing towards the North and South). Electrons and protons were simulated as coming from a 20$^\circ$ aperture cone at the same two azimuth and zenith positions. The extracted sub-layouts were analysed by two independent analysis packages: Eventdisplay from VERITAS, MAGIC Reconstruction Software (MARS). A cross-check for a small subset of configurations was also provided by the Image Pixel-wise fit for Atmospheric Cherenkov Telescopes (ImPACT) from H.E.S.S. (details can be found in \cite{APP_CTA_MC, evnDisplay, MARS, impact}).\\
As in the site selection case, the optimisation is performed with respect to the differential sensitivity, calculated as the average between the north and south pointing directions, of a point-like source located in the centre of the field of view. For simplicity, a single parameter is associated to each calculated sensitivity, the so-called performance per unit time (PPUT) \cite{Hassan2017Apj}, defined as: 
\begin{equation}
\mathrm{PPUT} =  \left( \prod_{i=1}^{N} \frac{F_\mathrm{sens,ref}(i)}{F_\mathrm{sens}(i)} \right)^{1/N}
\end{equation}
where $F_\mathrm{sens,ref}(i)$ and $F_\mathrm{sens}(i)$ are the reference and achieved sensitivity in the i-th bin in energy.\\
As described in section \ref{sec:intro}, the two main points for the baseline layout optimisation are full and sub-system performance. This will be done by optimising PPUTs both in the full energy range, and in the different sub-ranges relevant for each telescope type (see \cite{APP_CTA_MC,Hassan-2015}). By definition, higher PPUT values correspond to a better sensitivity in the considered energy range.\\

\section{Results}

In the following, the result of the layout optimisation will be presented for the Southern (Sec. \ref{south}) and the Northern (Sec. \ref{north}) Hemisphere sites.

\subsection{CTA-South baseline array}\label{south}

\begin{figure}[p]
\begin{center}
\centering\includegraphics[width=0.9\linewidth]{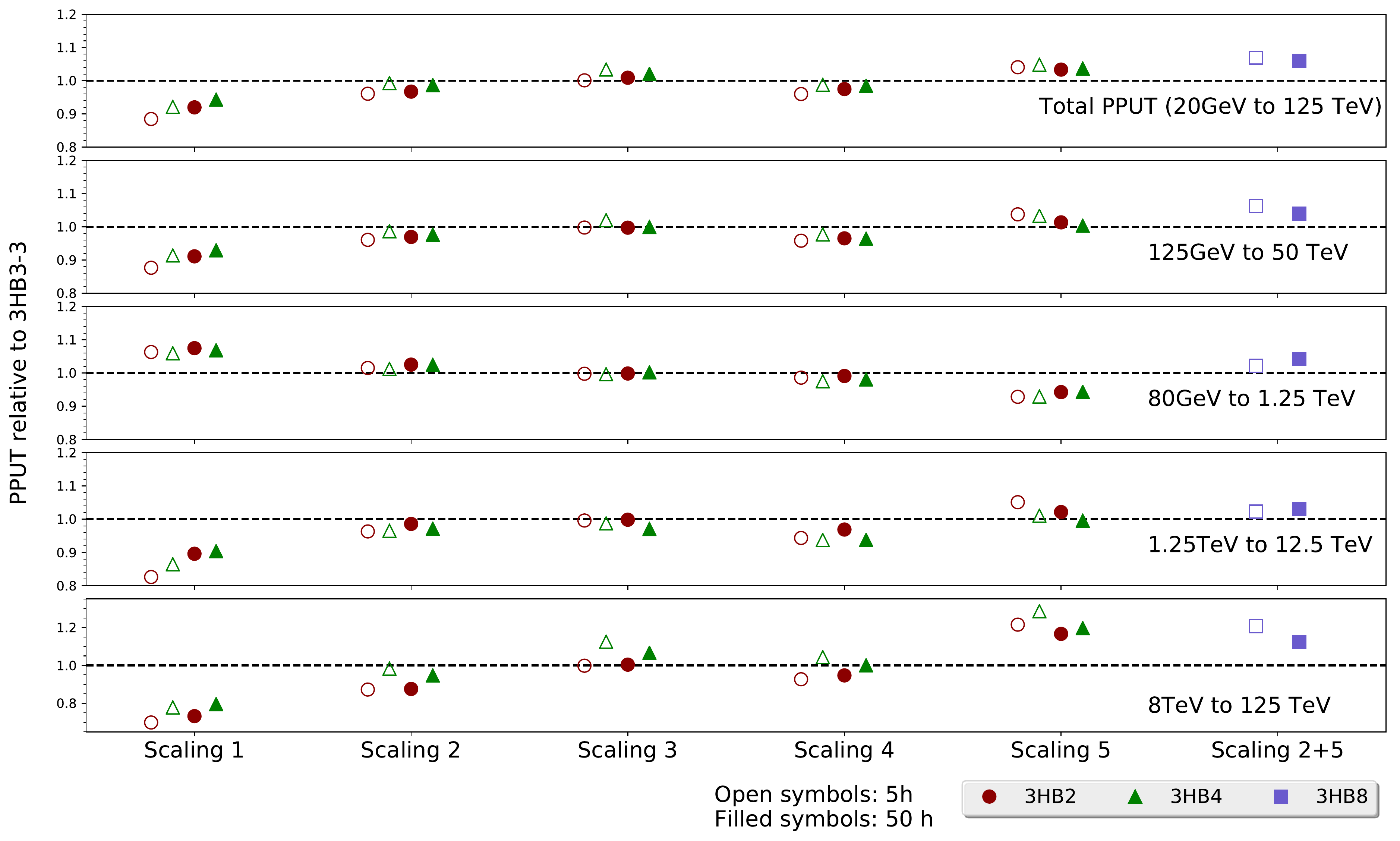}\\

\centering\includegraphics[width=0.9\linewidth]{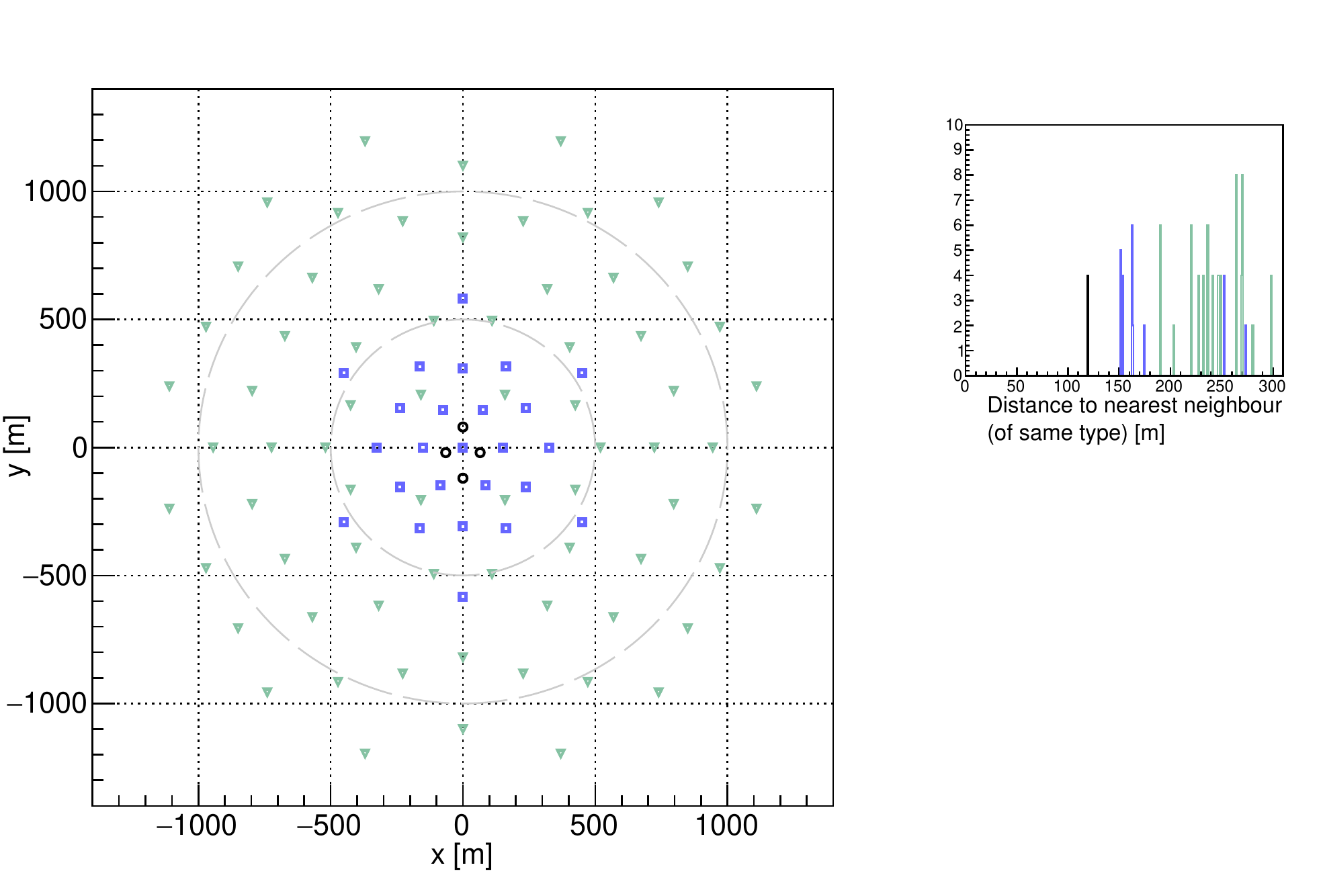}
\caption{
\textit{Top}: PPUTs relative to 3HB3-3 as a function of the scaling for different energy ranges for the Southern site. The trend shown here is not dependent on the layout used. ``Scaling 2+5'' refers to a layout obtained combining scaling 2 MSTs and scaling 5 SSTs. \textit{Bottom}: Layout of the proposed baseline for the Southern site. The LSTs are represented with black circles, MSTs with blue squares and SSTs with green triangles. On the bottom-right corner, the distance from the nearest neighbour of the same telescope type is presented using the same colour code of the left plot. Layout naming for Paranal (example: 3HB3-3) indicates the production version (3), the base pattern (H for hexagonal), the layout type (B for baseline), a variant number (in the example here: 3, as the one shown in Fig. \protect\ref{fig:distfactor}), and finally the scaling (here: 3).}
\label{fig:3HB9Layout}
\end{center}
\end{figure}

For the CTA Southern Hemisphere (CTA-S) layout, results on the scaling optimisation are shown in Fig. \ref{fig:3HB9Layout}, top panel. These results are irrespective of the choice of the specific layout used, and show the most efficient scaling both for the full array and each telescope type sub-system. Using this information, a new layout is defined by combining a mild scaling (scaling 2) for the MSTs and a strong scaling (scaling 5) for the SSTs. This combination provides optimal performance in the full energy range, maintaining competitive values with respect to other scalings for every sub-system.

Additional modifications were applied to define the final CTA-S baseline layout, shown in the bottom of Fig. \ref{fig:3HB9Layout}. Following the considerations listed in section \ref{sec:intro}, the performed optimisation takes mainly into account the full array and sub-array performances. In CTA-S, with no strong site physical constraints, staging and the impact on calibration and systematic uncertainties need to be taken into account.

Considering the staging for the LSTs, to which no scaling is applied, the proposed solution is an intermediate step between a square and a double-equilateral triangle. This configuration has the advantage that it performs significantly better than a square for a three LST stage, without a significant degradation of the full system performance. This compromise also works better than the double-triangle configuration for the situation where one of the east-west pair of telescopes is unavailable (for example due to maintenance activities). The east-west pair of telescopes has also a close to optimal performance for a two-telescope system.\\

To improve the MST and SST sub-systems calibration, while maintaining their stand-alone performance, small modifications were also applied to these layouts. For the MST layout, an additional telescope was added in the centre of the layout (surrounded by LSTs). This additional MST improves inter-telescope calibration between the LST and MST, and also removes the central gap of the MST stand-alone layout (shown, for example, in Fig. \ref{fig:distfactor} layouts). Following a similar reasoning, SST positions were also smoothed to populate internal regions of the full array, in which MSTs are present. This smoothing does not significantly affect neither the overall performance, nor the SST sub-system stand-alone performance.


\subsection{CTA-North baseline array}\label{north}

\begin{figure}[t]
\begin{center}
\centering\includegraphics[width=0.7\linewidth]{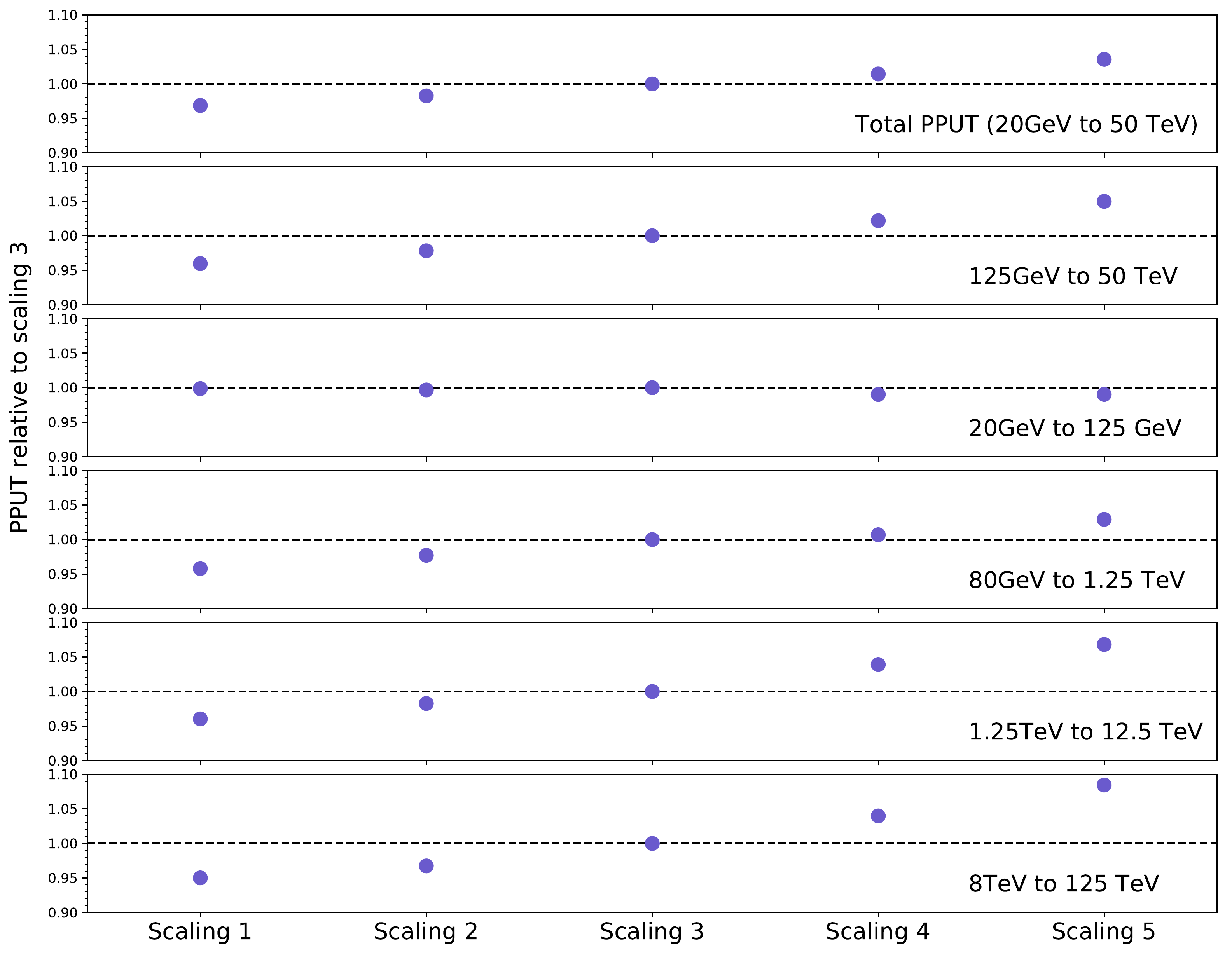}
\caption{PPUT ratio with respect to the PPUT of scaling 3, for different energy ranges and scaling factors for the La Palma site.}
\label{fig:PPUTLaPalma}
\end{center}
\end{figure}

\begin{figure}
\begin{center}
\includegraphics[width=0.49\textwidth]{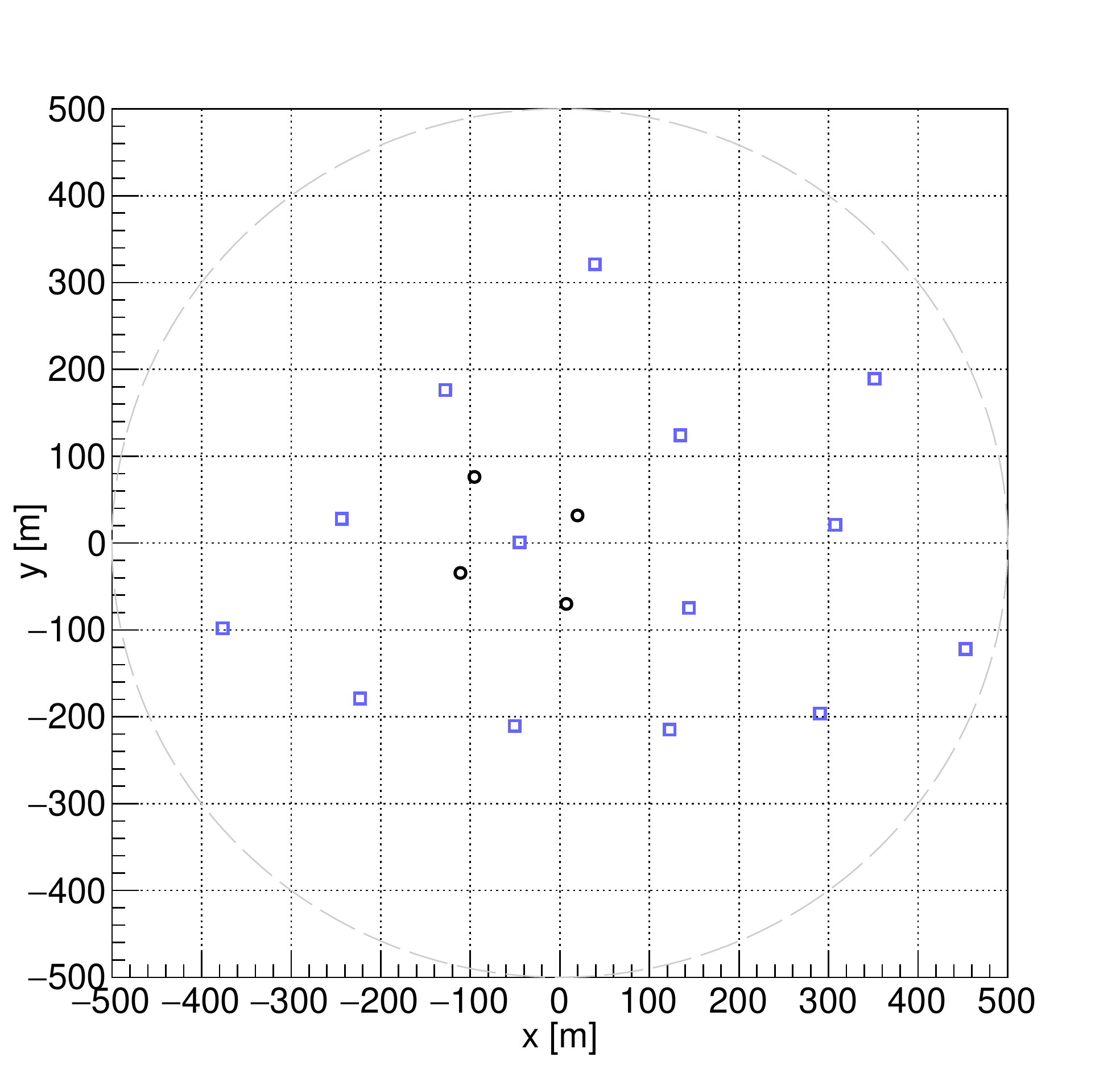}
\includegraphics[width=0.49\textwidth]{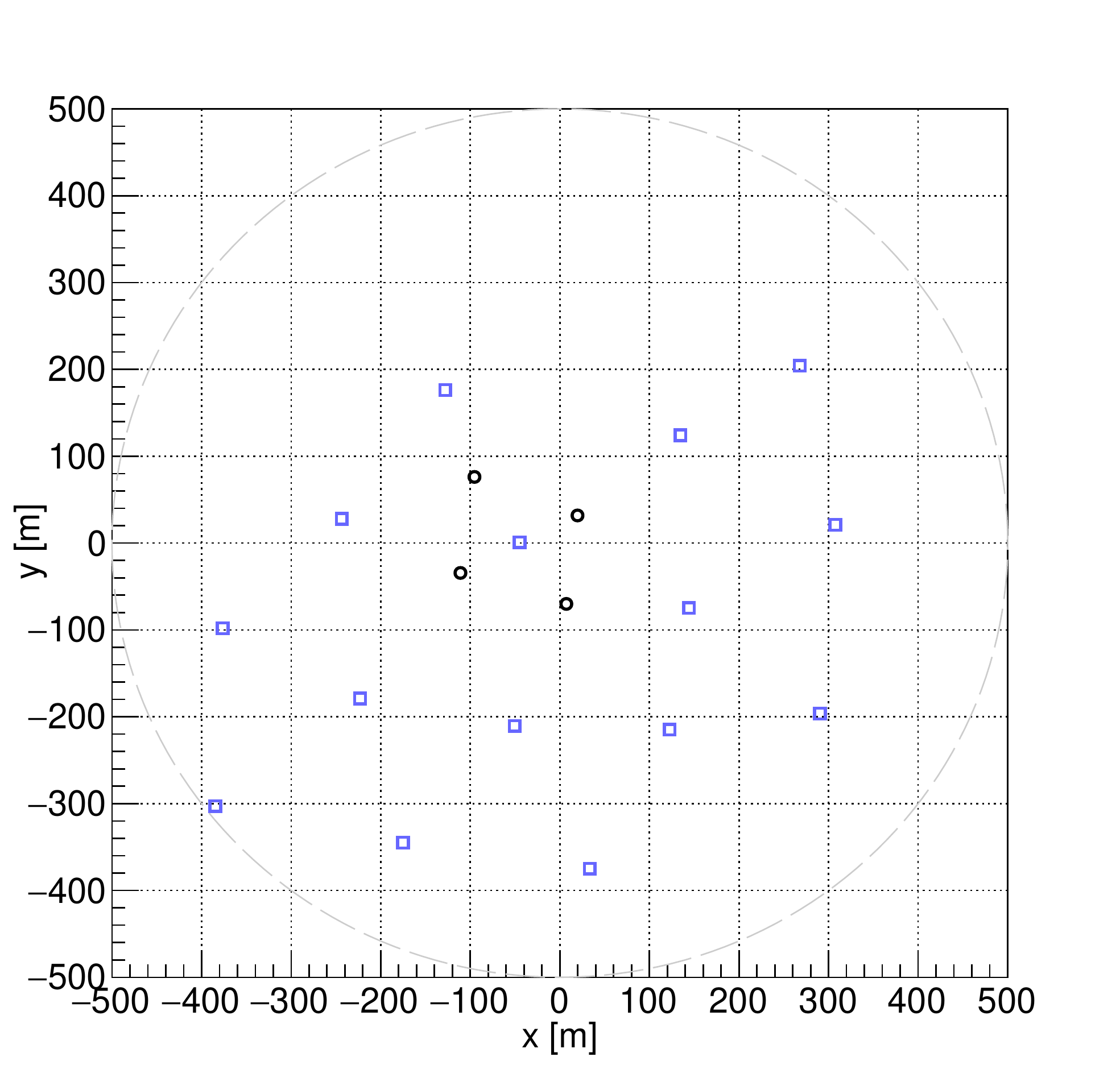}\\
\caption{Two layouts proposed as baseline arrays for the Northern site. They share the same LST positions (black circles) and roughly the same inter-telescope distances between MSTs (blue squares).}
\label{fig:AL4}
\end{center}
\end{figure}
For the CTA-North (CTA-N) array, composed of 4 LSTs and 19 MSTs, the best performance is obtained with the widest separation between the MSTs, $\sim 180$ m, as shown in Figure \ref{fig:PPUTLaPalma}. The observed trend does not depend on the sub-layout used, clearly showing that more extended MST layouts provide better sensitivity above $\sim$ 100 GeV, with no significant impact on performance within the lower energies. Larger separation between telescopes, while possible for few of them, are forbidden by the logistical constraints of the site.\\
While the positions of the LSTs are fixed, and the LST-1 is already under construction, more than one solution for the MST sub-layout is possible while maintaining the same inter-telescope distance, some of which are presented in Figure \ref{fig:AL4}. All these layouts are compliant with the constraints coming from the orography, roads and buildings and show a similar sensitivity. At the time of this study, it is therefore not possible to make a clear choice among these layouts, all providing excellent performance. The final CTA-N baseline layout will very likely be fixed once we attain a better understanding on the site constraints. 

\section{Conclusion}
The Cherenkov Telescope Array, the next generation instrument for Very High Energy gamma-ray astronomy, will be composed of two arrays, one per Earth Hemisphere. After the two sites have been selected \cite{Hassan2017Apj}, a new Monte Carlo production was carried out in order to define the best possible array layouts. With this purpose, several telescope positions were simulated with increasing distance among them.\\
The proposed final layouts are presented here. For the Northern array, where the telescope positions are severely constrained by external factors, several different layouts present comparable performance. For the Southern array a single layout, combination of mildly distanced MSTs and widely spaced SSTs, is proposed.\\
The final telescope positions may slightly differ from the baseline arrays presented here, as unforeseen external conditions might arise in the future causing the moving of one or more telescopes from the proposed configurations.

\subsubsection*{Acknowledgments}
We gratefully acknowledge financial support from the agencies and organizations listed here: http://www.cta-observatory.org/consortium\_acknowledgments

\bibliographystyle{JHEP}
\bibliography{references} %


\end{document}